\documentclass[aps,showpacs,preprintnumbers]{revtex4}

\usepackage{amsfonts}

\usepackage{graphicx}
\usepackage{dcolumn}
\usepackage{bm}
\usepackage{epsfig}
\newcommand{\vsig}{\mbox{\boldmath$\sigma$\unboldmath}}

\begin{document}

\title{The $K^-p\rightarrow \Sigma^0\pi^0$ reaction at low energies in a chiral quark model}
\author{
Xian-Hui Zhong$^{1,4}$ \footnote {E-mail: zhongxh@ihep.ac.cn} and
Qiang Zhao$^{2,3,4}$ \footnote {E-mail: zhaoq@ihep.ac.cn}}

\affiliation{ 1) Department of Physics, Hunan Normal University,
Changsha 410081, P.R. China }

\affiliation{ 2) Institute of High Energy Physics,
       Chinese Academy of Sciences, Beijing 100049, P.R. China
}
\affiliation{ 3) Department of Physics, University of Surrey,
Guildford, GU2 7XH, United Kingdom
            }
\affiliation{4) Theoretical Physics Center for Science Facilities,
Chinese Academy of Sciences, Beijing 100049, P.R. China}


\begin{abstract}

A chiral quark-model approach is extended to the study of the
$\bar{K}N$ scattering at low energies. The process of
$K^-p\rightarrow \Sigma^0\pi^0$ at $P_K\lesssim 800$ MeV/c (i.e. the
center mass energy $W\lesssim 1.7$ GeV) is investigated. This
approach is successful in describing the differential cross sections
and total cross section with the roles of the low-lying $\Lambda$
resonances in $n=1$ shells clarified. The $\Lambda(1405)S_{01}$
dominates the reactions over the energy region considered here.
Around $P_K\simeq 400$ MeV/c, the $\Lambda(1520)D_{03}$ is
responsible for a strong resonant peak in the cross section. The
$\Lambda(1670)S_{01}$ has obvious contributions around $P_K=750$
MeV/c, while the contribution of $\Lambda(1690)D_{03}$ is less
important in this energy region. The non-resonant background
contributions, i.e. $u$-channel and $t$-channel, also play important
roles in the explanation of the angular distributions due to
amplitude interferences. The $u$-channel turns out to have
significant destructive interferences with the $\Lambda(1405)S_{01}$
at the forward angles. In contrast, the $t$-channel $K^*$ exchange
has a constructive interference at the forward angles, while
suppresses the cross sections slightly at the backward angles. In
the $t$-channel, the $K^*$-exchange is more dominant over the
$\kappa$-exchange. Our analysis suggests that there exist
configuration mixings within the $\Lambda(1405)S_{01}$ and
$\Lambda(1670)S_{01}$ as admixtures of the
$[\textbf{70},^2\textbf{1},1/2]$ and
$[\textbf{70},^2\textbf{8},1/2]$ configurations. The
$\Lambda(1405)S_{01}$ is dominated by
$[\textbf{70},^2\textbf{1},1/2]$, and $\Lambda(1670)S_{01}$ by
$[\textbf{70},^2\textbf{8},1/2]$. The mixing angle is also
determined. The $\Lambda(1520)D_{03}$ and $\Lambda(1690)D_{03}$ are
assigned as the $[\textbf{70},^2\textbf{1},3/2]$ and
$[\textbf{70},^2\textbf{8},3/2]$, respectively.

\end{abstract}


\maketitle

\section{Introduction}


The reaction $K^-p\rightarrow \Sigma^0\pi^0$ is of particular
interest in the study of baryon resonances and $\bar{K}N$
interaction since there are no isospin-1 baryons contributing here
and it gives us a rather clean channel to study the $\Lambda$
resonances,  such as $\Lambda(1405)S_{01}$, $\Lambda(1670) S_{01}$,
$\Lambda(1520)D_{03}$ and $\Lambda(1690)D_{03}$.

In the literatures, many experimental
\cite{Manweiler:2008zz,armen:1970zh,London:1975av,Baxter:1974zs,
Mast:1974sx,Berley:1996zh,Bangerter:1980px,Starostin:2001zz,Prakhov:2004an,
Zychor:2007gf,Niiyama:2008rt,Zychor:2008ct} and theoretical efforts
\cite{Borasoy:2006sr,Dalitz:1959dn,Dalitz:1960du,Kaiser:1996js,Oset:1997it,Oller:2000fj,Pakvasa:1999zv,
Oset:2001cn,Jido:2003cb,GarciaRecio:2002td,Hyodo:2002pk,GarciaRecio:2003ks,
Magas:2005vu,Roca:2006kr,Roca:2006pu,Roca:2008kr,Sarkar:2005ap,Oller:2006jw,Oller:2006hx,Fink:1989uk,
Hyodo:2003qa,Borasoy:2005ie,Oller:2005ig,Hyodo:2003jw,Geng:2007hz,Geng:2007vm,Geng:2008er}
have been devoted to understanding  the nature of the low-lying
$\Lambda$ resonances. However, their properties still bare a lot of
controversies. For example, in the naive quark model the
$\Lambda(1405)$ is classified as the lowest $L=1$ orbital excited
$qqq$ state as an $SU(3)$ flavor singlet
\cite{Isgur78,Capstick:1986bm,Loring:2001ky}. Meanwhile, it is also
proposed to be a dynamically generated resonance emerging from the
interaction of the $\bar{K}N$ and $\pi \Sigma$ with a multi-quark
structure
\cite{Dalitz:1959dn,Dalitz:1960du,Kaiser:1996js,Oset:1997it,Oller:2000fj,
Oset:2001cn,Jido:2003cb,GarciaRecio:2002td,Hyodo:2002pk,GarciaRecio:2003ks,
Magas:2005vu,Roca:2008kr}. Most of those studies are based on the
unitary chiral perturbation theory (U$\chi$PT). Such a scenario is
developed further which proposes that the $\Lambda(1405)$ could be a
superposition of two resonances
\cite{Fink:1989uk,Hyodo:2003qa,Borasoy:2005ie,Oller:2005ig,Oller:2000fj,Jido:2003cb,Magas:2005vu}.
Similar mechanisms are studied in various processes
\cite{Hyodo:2003jw,Geng:2007hz,Geng:2007vm}, such as
$K^-p\rightarrow \pi^0\pi^0\Sigma^0$, $\pi^-p\rightarrow K^0\pi
\Sigma$ and $pp\rightarrow pK^+\Lambda$, as a support of the
dynamically generated states. How to clarify these issues and make a
contact with experimental observables are still an open
question~\cite{Zychor:2007gf,Zychor:2008ct,Zou:2007mk,Xie:2007qt,Sibirtsev:2005mv}.

On the other hand, it is of great importance to understand the
excitation of those low-lying $\Lambda$ states in a quark model
framework. Quark model somehow provides a guidance for the
underlying effective degrees of freedom within hadrons. In order to
probe exotic configurations such as multiquarks and hybrids, one
should also have a good understanding of where the non-relativistic
constituent quark model (NRCQM) breaks down. Particularly in the
sector of hyperon states, there are still a lot of ambiguities to be
clarified. Apart from the $\Lambda(1405)$, the $\Lambda(1520)$ and
$\Lambda(1670)$ are also suggested to be quasibound states of a
meson and a baryon, which are dynamically generated resonances
 based on the U$\chi$PT~\cite{Sarkar:2005ap,Roca:2008kr,Roca:2006kr}.
While in the quark model these two states are classified as the
lowest $L=1$ orbital excited states with $J^P=3/2^-$ and
$J^P=1/2^-$, respectively. To clarify the nature of those low-lying
$\Lambda$ resonances and their internal effective quark degrees of
freedom, more theoretical and experimental studies are needed.

Recently, the higher precision data of the reaction $K^-p\rightarrow
\Sigma^0\pi^0$ at eight momentum beams between 514 and 750 MeV/c
were reported \cite{Manweiler:2008zz}, which provides us a good
opportunity to study the properties of these low-lying $\Lambda$
resonances. In this work, we make an investigation of the
$K^-p\rightarrow \Sigma^0\pi^0$ reaction in a chiral quark model. In
this model an effective chiral Lagrangian is introduced to account
for the quark-pseudoscalar-meson coupling. Since the quark-meson
coupling is invariant under the chiral transformation, some of the
low-energy properties of QCD are retained. The chiral quark model
has been well developed and widely applied to meson photoproduction
reactions~\cite{qk1,qk2,qkk,Li:1997gda,zhao-kstar,qk3,qk4,qk5,He:2008ty}.
Its recent extension to describe the process of $\pi N$ scattering
~\cite{Zhong:2007fx} and investigate the strong decays of charmed
hadrons ~\cite{Zhong:2007gp,Zhong:2008kd} also turns out to be
successful and inspiring.

In the literatures the $\bar{K}N$ scattering has been studied using
different approaches, such as the $K$-matrix methods
\cite{Martin:1969ud}, dispersion relations
\cite{Gensini:1997fp,Martin:1980qe}, meson-exchange models
\cite{Buttgen:1985yz,Buettgen:1990yw,MuellerGroeling:1990cw},
coupled-channel approaches \cite{Borasoy:2006sr,Oller:2000fj,
Oset:1997it,Oset:2001cn,GarciaRecio:2002td,Hyodo:2003qa,Hyodo:2002pk},
and quark models \cite{Hamaie:1995wy}. Compared with these models,
our model has several obvious features. One is that only a limited
number of parameters will appear in the formalism. In particular,
only one parameter is need for the resonances to be coupled to the
pseudoscalar meson. This distinguishes from hadronic models where
each resonance requires one additional coupling constant as a free
parameter. The second is that all the resonances can be treated
consistently at quark level. Thus, it has predictive powers when
exposed to experimental data, and information about the resonance
structures and form factors can be extracted.

In the $K^-p\rightarrow \Sigma^0\pi^0$ reaction, for the
$s$-channel, the $K^-$- and the $\pi^0$-mesons can not couple to the
same quark in a baryon, which leads to a strong suppression in the
$s$-channel amplitudes. As shown in Fig.~\ref{fig-1}, the amplitude
$M_2^s$ is suppressed relative to $M_3^s$ by a factor of $(-1/2)^n$
with $n$ for the main quantum number of the NRCQM harmonic
oscillator
potential~\cite{qk1,qk2,qkk,Li:1997gda,zhao-kstar,qk3,qk4,qk5,Zhong:2007fx,He:2008ty}.
In contrast, it is allowed for the $u$-channel that the kaon and
pion are coupled to the same quark. Thus, the $u$-channel gives a
large background in the cross section, and has significant
destructive interferences with $\Lambda(1405)S_{01}$ at the forward
angles. The $t$-channel, dominated by the $K^*$ exchange, also plays
an important role in the reactions. It suppresses the cross section
obviously at the backward angles, while enhances it at the forward
angles. We also consider the $t$-channel scalar meson exchange, i.e.
$\kappa$, but find its contributions are negligibly small.

The $\Lambda(1405)$ governs the reaction in the whole energy region
near threshold which is similar to the $S_{11}(1535)$ dominance in
$\pi^- p\to \eta n$~\cite{Zhong:2007fx}. Around $P_K= 400$ MeV/c,
the $\Lambda(1520)$ is responsible for the sharp resonant peak in
the total cross section. The contributions of $\Lambda(1670)$ turn
out to be important at $P_K\simeq 750$ MeV/c.

The paper is organized as follows. In the subsequent section, the
amplitudes of $s$- and $u$-channels are obtained. Then, amplitudes
of $t$-channel are given in Sec.\ \ref{tc}. The resonance
contributions are separated in Sec.\ \ref{ss}. We present our
calculations and discussions in Sec.\ \ref{cy}. Finally, a summary
is given in Sec.\ \ref{sum}.

\section{amplitudes of the $s$- and $u$-channel transitions}\label{suc}

\subsection{The interactions }\label{qmc}

The effective quark-pseudoscalar-meson coupling in the chiral quark
model has been discussed in detail in
Refs.~\cite{Li:1997gda,qk3,Zhong:2007fx}. Here, we only outline the
main formulae to keep the self-consistence of this work.

The low energy quark-meson interactions are described by the
effective Lagrangian \cite{Li:1997gda,qk3}
\begin{eqnarray} \label{lg}
\mathcal{L}=\bar{\psi}[\gamma_{\mu}(i\partial^{\mu}+V^{\mu}+\gamma_5A^{\mu})-m]\psi
+\cdot\cdot\cdot,
\end{eqnarray}
where $V^{\mu}$ and $A^{\mu}$ correspond to vector and axial
currents, respectively. They are given by
\begin{eqnarray}
V^{\mu} &=&
 \frac{1}{2}(\xi\partial^{\mu}\xi^{\dag}+\xi^{\dag}\partial^{\mu}\xi),
\nonumber\\
 A^{\mu}
&=&
 \frac{1}{2i}(\xi\partial^{\mu}\xi^{\dag}-\xi^{\dag}\partial^{\mu}\xi),
\end{eqnarray}
under the chiral transformation $\xi=\exp{(i \phi_m/f_m)}$, where
$f_m$ is the meson's decay constant. For the $SU(3)$ case, the
pseudoscalar-meson octet $\phi_m$ can be expressed as
\begin{eqnarray}
\phi_m=\pmatrix{
 \frac{1}{\sqrt{2}}\pi^0+\frac{1}{\sqrt{6}}\eta & \pi^+ & K^+ \cr
 \pi^- & -\frac{1}{\sqrt{2}}\pi^0+\frac{1}{\sqrt{6}}\eta & K^0 \cr
 K^- & \bar{K}^0 & -\sqrt{\frac{2}{3}}\eta},
\end{eqnarray}
and the quark field $\psi$ is given by
\begin{eqnarray}\label{qf}
\psi=\pmatrix{\psi(u)\cr \psi(d) \cr \psi(s) }.
\end{eqnarray}

At the leading order of the Lagrangian [Eq.(\ref{lg})], the
quark-meson pseudovector coupling is
\begin{eqnarray}\label{coup}
H_m=\sum_j
\frac{1}{f_m}\bar{\psi}_j\gamma^{j}_{\mu}\gamma^{j}_{5}\psi_j\vec{\tau}\cdot\partial^{\mu}\vec{\phi}_m.
\end{eqnarray}
where $\psi_j$ represents the $j$-th quark field in a hadron.

The non-relativistic form of Eq. (\ref{coup}) can be written as
\cite{Zhong:2007fx,Li:1997gda,qk3}
\begin{eqnarray}\label{ccpk}
H^{nr}_{m}=\sum_j\Big\{\frac{\omega_m}{E_f+M_f}\vsig_j\cdot
\textbf{P}_f+ \frac{\omega_m}{E_i+M_i}\vsig_j \cdot
\textbf{P}_i  
-\vsig_j \cdot \textbf{q} +\frac{\omega_m}{2\mu_q}\vsig_j\cdot
\textbf{p}'_j\Big\}I_j \varphi_m,
\end{eqnarray}
where $\vsig_j$ corresponds to the Pauli spin vector of the $j$-th
quark in a hadron, and $\mu_q$ is a reduced mass given by
$1/\mu_q=1/m_j+1/m'_j$, where $m_j$ and $m'_j$ stand for the masses
of the $j$-th quark in the initial and final hadrons, respectively.
For emitting a meson, we have $\varphi_m=\exp({-i\textbf{q}\cdot
\textbf{r}_j})$, and for absorbing a meson we have
$\varphi_m=\exp({i\textbf{q}\cdot \textbf{r}_j})$. In the above
non-relativistic expansions, $\textbf{p}'_j$
$(=\textbf{p}_j-\frac{m_j}{M}\mathbf{P}_{c.m.})$ is the internal
momentum for the $j$-th quark in the initial meson rest frame.
$\omega_m$ and $\textbf{q}$ are the energy and three-vector momentum
of the light meson, respectively. The isospin operator $I_j$ in Eq.
(\ref{ccpk}) is expressed as
\begin{eqnarray}
I_j=\cases{ a^{\dagger}_j(u)a_j(s) & for $K^+$, \cr
a^{\dagger}_j(s)a_j(u) & for $K^-$,\cr a^{\dagger}_j(d)a_j(s) & for
$K^0$, \cr a^{\dagger}_j(s)a_j(d) & for $\bar{K^0}$,\cr
a^{\dagger}_j(u)a_j(d) & for $\pi^+$,\cr a^{\dagger}_j(d)a_j(u)  &
for $\pi^-$,\cr
\frac{1}{\sqrt{2}}[a^{\dagger}_j(u)a_j(u)-a^{\dagger}_j(d)a_j(d)] &
for $\pi^0$,}
\end{eqnarray}
where $a^{\dagger}_j(u,d,s)$ and $a_j(u,d,s)$ are the creation and
annihilation operators for the $u$, $d$ and $s$ quarks.

\begin{center}
\begin{figure}[ht]
\centering \epsfxsize=9 cm \epsfbox{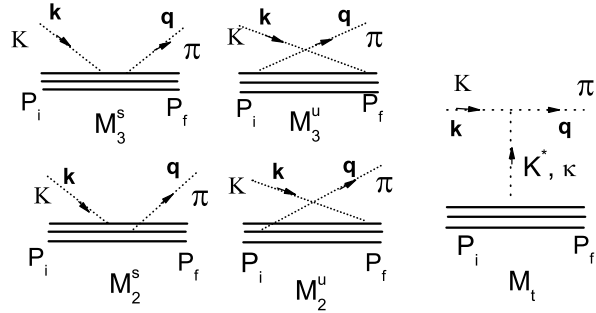} \caption{ Transition
channels labeled by the Mandelstem variables, i.e. $s$, $u$, and
$t$- channels. $M^s_3$ and $M^u_3$ ($M^s_2$, $M^u_2$ ) correspond to
the amplitudes of the $s$- and $u$-channels with the incoming meson
and outgoing meson absorbed and emitted by the same quark (different
quarks), respectively. Note that in reaction $K^- p\to
\Sigma^0\pi^0$ the amplitude $M^s_3$ vanishes.}\label{abc}
\end{figure}
\end{center}

\subsection{The $s$-channel amplitudes} \label{s}

The $s$-channel transition amplitudes as shown in Fig. \ref{abc} can
be expressed as
\begin{eqnarray}
\mathcal{M}_{s}=\sum_j\langle N_f |H_{\pi} |N_j\rangle\langle N_j
|\frac{1}{E_i+\omega_K-E_j}H_{K }|N_i\rangle, \label{sc}
\end{eqnarray}
where $\omega_K$ is the energy of the incoming $K^-$-meson. $H_K$
and $H_\pi$ are the standard quark-meson couplings at tree level
described by Eq.(\ref{coup}). $|N_i\rangle$, $|N_j\rangle$ and
$|N_f\rangle$ stand for the initial, intermediate and final states,
respectively, and their corresponding energies are $E_i$, $E_j$ and
$E_f$, which are the eigenvalues of the NRCQM Hamiltonian
$\hat{H}$~\cite{Isgur78, Isgur:1977ky}. Following the procedures
developed in Refs. \cite{qkk,Li:1997gda,qk3,Zhong:2007fx}, one can
then express the $s$-channel amplitudes by operator expansions:
\begin{eqnarray} \mathcal{M}_{s}=\sum_j\langle N_f
|H_{\pi} |N_j\rangle\langle N_j| \sum_n
\frac{1}{\omega_K^{n+1}}(\hat{H}-E_i)^n H_{K }|N_i\rangle \ ,
\end{eqnarray}
where $n$ is the principle harmonic oscillator quantum number. Note
that for any operator ${\hat{\cal O}}$, one has
\begin{eqnarray} (\hat{H}-E_i){\hat{\cal
O}}|N_i\rangle = [\hat{H}, \ {\hat{\cal O}}]|N_i\rangle ,
\end{eqnarray}
a systematic expansion of the commutator between the NRCQM
Hamiltonian $\hat{H}$ and the vertex coupling $H_K$ and $H_\pi$ can
thus be carried out. Details of this treatment can be found in Refs.
\cite{qkk,Li:1997gda,qk3}, but we note that in this study only the
spin-independent potential in $\hat{H}$ is considered as a feasible
leading order calculation.

Finally, we can obtain the $s$-channel amplitude in the harmonic
oscillator basis, which is expressed as \cite{Zhong:2007fx}
\begin{eqnarray}
\mathcal{M}^{s}=\sum_n
(\mathcal{M}^{s}_{3}+\mathcal{M}^{s}_{2})e^{-(\textbf{k}^2+\textbf{q}^2)/6\alpha^2},
\end{eqnarray}
where $\alpha$ is the oscillator strength, and
$e^{-(\textbf{k}^2+\textbf{q}^2)/6\alpha^2}$ is a form factor in the
harmonic oscillator basis. $\mathcal{M}^{s}_{3}$
($\mathcal{M}^{s}_{2}$) corresponds to the amplitudes for the
outgoing meson and incoming meson absorbed and emitted by the same
quark (different quarks) (see Fig. \ref{abc}). Because of the
isospin selection rule, the $\pi$ and $K^-$ can not couple to the
same quark. Thus, the contribution of $\mathcal{M}^{s}_{3}$ vanishes
and only $\mathcal{M}^{s}_{2}$ contributes to the $s$-channel, i.e.
\begin{eqnarray}\label{ms2}
\mathcal{M}^{s}_{2}&=&\langle N_f |6I_1\Big\{\vsig_{1}\cdot
\textbf{A}_{out}\vsig_3 \cdot \textbf{A}_{in}\sum_{n=0}
\frac{F_s(n)}{n !}\frac{\mathcal{X}^n}{(-2)^n} +\Big[-
\vsig_{1}\cdot \textbf{A}_{out}\frac{\omega_{K}}{6\mu_q}\vsig_3
\cdot \textbf{q}-\frac{\omega_{\pi}}{3m_q}\vsig_1 \cdot
\textbf{k}\vsig_{3}\cdot \textbf{A}_{in} \nonumber\\
&&+ \frac{\omega_{\pi}}{m_q}\frac{\omega_{K}}{2\mu_q}
\frac{\alpha^2}{3}\vsig_1\cdot\vsig_3\Big]\times\sum_{n=1}
\frac{F_s(n)}{(n-1) !}\frac{\mathcal{X}^{n-1}}{(-2)^n}+
\frac{\omega_{\pi}}{3m_q}\frac{\omega_{K}}{6\mu_q}\vsig_1 \cdot
\textbf{q}\vsig_{3}\cdot \textbf{k}\sum_{n=2} \frac{F_s(n)}{(n-2)
!}\frac{\mathcal{X}^{n-2}}{(-2)^{n}}\Big\}|N_i\rangle,
\end{eqnarray}
with
\begin{eqnarray}
\mathbf{A}_{in}=-\left(1+ \omega_{K} \mathcal{K}_i-\frac{\omega_K}{6\mu_q}\right)\textbf{k},\\
\mathbf{A}_{out}=-\left(1+\omega_{\pi}\mathcal{K}_f-\frac{\omega_\pi}{3m_q}\right)\textbf{q},
\end{eqnarray}
where $\mathcal{K}_i=1/(E_i+M_i)$, $\mathcal{K}_f=1/(E_f+M_f)$ and
$m_q$ is the light quark mass. In Eq. (\ref{ms2}), the subscriptions
of the spin operator $\vsig$ denote that it either operates on quark
3 or quark 1. The $\mathcal{X}$ is defined as
$\mathcal{X}\equiv\frac{\textbf{k}\cdot \textbf{q}}{3 \alpha^2}$,
and the factor $F_s(n)$ is given by expanding the energy propagator
in Eq. (\ref{sc})  which leads to
\begin{eqnarray}\label{s-prop}
F_s(n)=\frac{M_n}{P_i \cdot k-nM_n \omega_h},
\end{eqnarray}
where $M_n$ denotes the mass of the excited state in the $n$-th
shell, while $\omega_h$ is the typical energy of the harmonic
oscillator; $P_i$ and $k$ are the four momenta of the initial state
nucleon and incoming $K^-$ meson in the c.m. system. The $F_s(n)$
has clear physical meaning that recovers the hadronic level
propagators. We will come back to this in the next section.

The above transition amplitude can be written coherently in terms of
a number of $g$-factors, which will allow us to relate the
quark-level amplitudes to those at hadronic level
\begin{eqnarray} \label{sac}
\mathcal{M}_{s}&=&\Big\{g_{s2}\textbf{A}_{out}\cdot\textbf{A}_{in}\sum_{n=0}
(-2)^{-n} \frac{F_s(n)}{n !}\mathcal{X}^n
+g_{s2}\left(-\frac{\omega_{K}}{6\mu_q}\textbf{A}_{out}\cdot
\textbf{q}-\frac{\omega_{\pi}}{3m_q}\textbf{A}_{in}\cdot
\textbf{k}+\frac{\omega_{\pi}}{m_q}\frac{\omega_{K}}{2\mu_q}
\frac{\alpha^2}{3}\right)\nonumber\\
&&\times\sum_{n=1}(-2)^{-n}\frac{F_s(n)}{(n-1) !}\mathcal{X}^{n-1}+
g_{s2}\frac{\omega_{\pi}\omega_{K}}{18m_q\mu_q}\textbf{k}\cdot\textbf{q}
\sum_{n=2}\frac{F_s(n)}{(n-2)!}(-2)^{-n}\mathcal{X}^{n-2}\nonumber\\
&&+g_{v2}i\vsig
\cdot(\textbf{A}_{out}\times\textbf{A}_{in})\sum_{n=0}(-2)^{-n}
\frac{F_s(n)}{n !}\mathcal{X}^n
+g_{v2}\frac{\omega_{\pi}\omega_{K}}{18m_q\mu_q}i\vsig\cdot(\textbf{q}\times\textbf{k})
\nonumber\\
&&
\times\sum_{n=2}(-2)^{-n}\frac{F_s(n)}{(n-2)!}\mathcal{X}^{n-2}\Big\}e^{-(\textbf{k}^2+\textbf{q}^2)/6\alpha^2},
\end{eqnarray}
where the $g$-factors, $g_{s2}$ and $g_{v2}$, in the $s$-channel are
defined as
\begin{eqnarray}
g_{s2}&\equiv&\langle N_f |\sum_{i\neq j} I^{\pi}_i
I^{K}_j\vsig_i\cdot \vsig_j |N_i
\rangle/3,\\
g_{v2}&\equiv&\langle N_f |\sum_{i\neq j} I^{\pi}_i
I^{K}_j(\vsig_i\times \vsig_j)_z |N_i\rangle/2,
\end{eqnarray}
which can be derived from the quark model in the $SU(6)\otimes O(3)$
limit.

\subsection{The $u$-channel amplitudes}

The $u$-channel transition amplitudes (see Fig. \ref{abc}) are given
by
\begin{eqnarray}
\mathcal{M}_{u}=\sum_j\langle N_f |H_{K }
\frac{1}{E_i-\omega_\pi-E_j}|N_j\rangle\langle N_j | H_{\pi }
|N_i\rangle, \label{uc}
\end{eqnarray}
Following the same procedure in \ref{s}, when the outgoing  and
incoming mesons couple to the same quark, we obtain the amplitude
\begin{eqnarray}\label{mu3}
\mathcal{M}^{u}_{3}&=&-\langle N_f |3I^K_3
I^{\pi}_3\Big\{\vsig_{3}\cdot \textbf{B}_{in}\vsig_3 \cdot
\textbf{B}_{out}\sum_{n=0} F_u(n)\frac{1}{n !}\mathcal{X}^n +
\Big[-\vsig_{3}\cdot \textbf{B}_{in}\frac{\omega_{\pi}}{3m_q}\vsig_3
\cdot \textbf{k} - \frac{\omega_{K}}{6\mu_q}\vsig_3 \cdot
\textbf{q}\vsig_{3}\cdot \textbf{B}_{out}+
\frac{\omega_{\pi}}{m_q}\frac{\omega_{K}}{2\mu_q}
\frac{\alpha^2}{3}\Big]\nonumber\\
&& \times\sum_{n=1} F_u(n)\frac{\mathcal{X}^{n-1}}{(n-1) !}+
\frac{\omega_{\pi}}{3m_q}\frac{\omega_{K}}{6\mu_q}\vsig_3 \cdot
\textbf{k}\vsig_{3}\cdot \textbf{q}\sum_{n=2}
F_u(n)\frac{\mathcal{X}^{n-2}}{(n-2) !}\Big\}|N_i\rangle.
\end{eqnarray}
While the outgoing  and incoming mesons couple to two different
quarks, the transition amplitude is given by
\begin{eqnarray}\label{mu2}
\mathcal{M}^{u}_{2}&=&-\langle N_f
|6I^K_1I^{\pi}_3\Big\{\vsig_{1}\cdot \textbf{B}_{in}\vsig_3 \cdot
\textbf{B}_{out}\sum_{n=0} \frac{F_u(n)}{n !}\frac{\mathcal{X}^n
}{(-2)^n}+\Big[ -\vsig_{1}\cdot
\textbf{B}_{in}\frac{\omega_{\pi}}{3m_q}\vsig_3 \cdot
\textbf{k}-\frac{\omega_{K}}{6\mu_q}\vsig_1 \cdot
\textbf{q}\vsig_{3}\cdot \textbf{B}_{out}\nonumber\\&&+
\frac{\omega_{\pi}}{m_q} \frac{\omega_{K}}{2\mu_q}
\frac{\alpha^2}{3}\vsig_1 \cdot \vsig_3\Big] \sum_{n=1}
\frac{F_u(n)}{(n-1) !}\frac{\mathcal{X}^{n-1}}{(-2)^n}+
\frac{\omega_{\pi}}{3m_q}\frac{\omega_{K}}{6\mu_q}\vsig_1 \cdot
\textbf{k}\vsig_{3}\cdot \textbf{q}\sum_{n=2} \frac{F_u(n)}{(n-2)
!}\frac{\mathcal{X}^{n-2}}{(-2)^{n}}\Big\}|N_i\rangle .
\end{eqnarray}
In the above equations, we have defined
\begin{eqnarray}
\mathbf{B}_{in}\equiv-\omega_K\left(\mathcal{K}_f+\mathcal{K}_j-\frac{1}{6\mu_q}\right)\mathbf{q}-(1+\omega_K\mathcal{K}_j)\mathbf{k},\\
\mathbf{B}_{out}\equiv-\omega_\pi\left(\mathcal{K}_i+\mathcal{K}_j-\frac{1}{3m_q}\right)\mathbf{k}-(1+\omega_\pi
\mathcal{K}_j)\mathbf{q},
\end{eqnarray}
where $\mathcal{K}_j=1/(E_j+M_j)$.

In Eqs.(\ref{mu3}) and (\ref{mu2}), the factor $F_u(n)$ is written
as
\begin{eqnarray}
F_u(n)=\frac{M_n}{P_i \cdot q+nM_n \omega_h},
\end{eqnarray}
where $q$ is the four momentum of the outgoing $\pi$ meson in the
c.m. system.

The total amplitude for the $u$-channel is expressed as
\begin{eqnarray}\label{uac}
\mathcal{M}_{u}&=&-\Big\{\textbf{B}_{in}\cdot\textbf{B}_{out}\sum_{n=0}
\left[g^u_{s1}+(-2)^{-n}g^u_{s2}\right] \frac{F_u(n)}{n
!}\mathcal{X}^n
+\left(-\frac{\omega_{\pi}}{3m_q}\textbf{B}_{in}\cdot
\textbf{k}-\frac{\omega_{K}}{3m_q}\textbf{B}_{out}\cdot
\textbf{q}+\frac{\omega_{K}}{2\mu_q}\frac{\omega_{\pi}}{m_q}
\frac{\alpha^2}{3}\right)\nonumber\\
&&\times\sum_{n=1}[g^u_{s1}+(-2)^{-n}g^u_{s2}] \frac{F_u(n)}{(n-1)
!}\mathcal{X}^{n-1}+
\frac{\omega_{\pi}\omega_{K}}{18m_q\mu_q}\textbf{k}\cdot\textbf{q}
\sum_{n=2}\frac{F_u(n)}{(n-2)!}[g^u_{s1}+(-2)^{-n}g^u_{s2}]
\mathcal{X}^{n-2} \nonumber\\
&& + i\vsig
\cdot(\textbf{B}_{in}\times\textbf{B}_{out})\sum_{n=0}\left[g^u_{v1}+(-2)^{-n}g^u_{v2}\right]
\frac{F_u(n)}{n !}\mathcal{X}^n
-\frac{\omega_{\pi}\omega_{K}}{18m_q\mu_q}i\vsig\cdot(\textbf{q}\times\textbf{k})
\sum_{n=2}[g^u_{v1}+(-2)^{-n}g^u_{v2}]\nonumber\\
&&\times\frac{F_u(n)}{(n-2)!}\mathcal{X}^{n-2} +i\vsig \cdot
\left[-\frac{\omega_{\pi}}{3m_q}(\textbf{B}_{in}\times
\textbf{k})-\frac{\omega_{K}}{6\mu_q}(\textbf{q}\times\textbf{B}_{out}
)\right ]\sum_{n=1}\left[g^u_{v1}+(-2)^{-n}g^u_{v2}\right]
\mathcal{X}^{n-1}
\frac{F_u(n)}{(n-1) !} \Big\}\nonumber\\
&&\times e^{-(\textbf{k}^2+\textbf{q}^2)/6\alpha^2},
\end{eqnarray}
where the $g$ factors in the $u$-channel are determined by
\begin{eqnarray}
g^u_{s1}&\equiv&\langle N_f |\sum_{j}I^{K}_j I^{\pi}_j |N_i \rangle,\\
g^u_{s2}&\equiv&\langle N_f |\sum_{i\neq j} I^{K}_i
I^{\pi}_j\vsig_i\cdot \vsig_j |N_i
\rangle/3,\\
g^u_{v1}&\equiv&\langle N_f |\sum_{j}
I^{K}_j I^{\pi}_j \sigma^z_j |N_i\rangle,\\
g^u_{v2}&\equiv&\langle N_f |\sum_{i\neq j} I^{K}_i I^{\pi}_j
(\vsig_i\times \vsig_j)_z |N_i\rangle/2.
\end{eqnarray}
The numerical values of these factors can be derived in the
$SU(6)\otimes O(3)$ symmetry limit.

The first term in Eqs. (\ref{ms2}), (\ref{mu3}) and (\ref{mu2})
comes from the correlation between the c.m. motion of the $K^-$
meson transition operator and the c.m. motion of $\pi$-meson
transition operator; the second and the third terms are the
correlation among the internal and the c.m. motions of the $K^-$ and
$\pi$ transition operators, and their contributions begin with the
$n\geq 1$ exited states in the harmonic oscillator basis. The last
two terms in these equations correspond to the correlation of the
internal motions between the $K^-$ and $\pi$ transition operators,
and their contributions begin with either $n\geq 1$ or $n\geq 2$
exited states.

\section{Amplitudes of the $t$-channel transitions}\label{tc}

\subsection{The interactions}

The light meson exchange in the $t$-channel at low energies will
generally have larger contributions than the heavy ones. In $K^-
p\to \Sigma^0\pi^0$, we consider the $t$-channel vector meson
$K^*(892)$ and scalar meson $\kappa(800)$ exchanges which are found
dominantly coupled to $K\pi$~\cite{PDG}.


For the $K^*K\pi$ and $\kappa K\pi$ couplings, we introduce the
following effective interactions
\begin{eqnarray}
H_{K^*K\pi}&=&iG_v\{[(\partial_\mu
\bar{K})K^*-\bar{K}^*(\partial_\mu K)]\vec{\tau}\cdot\vec{\pi}
-[\bar{K}K^*-\bar{K}^*K]\vec{\tau}\cdot (\partial_\mu
\vec{\pi})\},\\
H_{\kappa K\pi}&=&\frac{g_{\kappa K\pi}}{2m_\pi}\partial_\mu
K\partial^\mu \pi \kappa,
\end{eqnarray}
where $G_v$ and $g_{\kappa K\pi}$ are the coupling constants to be
determined by experimental data~\cite{PDG}.

Similar to the quark-pseudoscalar-meson coupling, we introduce the
$K^*NN$ and $\kappa NN$ couplings at quark level by effective
$K^*qq$ and $\kappa qq$ Lagrangians:
\begin{eqnarray}
H_{K^*qq}&=& \bar{\psi}_j(a\gamma^{\nu}+\frac{i
b\sigma^{\nu\lambda}q_{\lambda}}{2m_q})K^{*}_{\nu}
\psi_j,\\
H_{\kappa qq}&=& g_{\kappa qq}\bar{\psi}_j\psi_j
\kappa\label{acoup},
\end{eqnarray}
where the constants $a$, $b$ and $g_{\kappa qq}$ are the vector,
tensor and scalar coupling constants, which are treated as free
parameters in this work.

\subsection{The amplitudes}

For the vector meson $K^*$-exchange, the amplitude of $t$-channel
can be written as
\begin{eqnarray} \label{tchannel}
\mathcal{M}^V_t= G_v(q^{\mu}+k^\mu)G_{\mu\nu}\sum_j
\bar{\psi}_j(a\gamma^{\nu}+\frac{i
b\sigma^{\nu\lambda}q_{\lambda}}{2m_q})\phi^{m}_{\nu} \psi_j,
\end{eqnarray}
where  $q^{\mu}, k^\mu$ are the four momenta of the $\pi^0$ and
$K^-$ mesons, respectively. In (\ref{tchannel}), the propagator
$G_{\mu\nu}$ is defined by
\begin{eqnarray}
G_{\mu\nu}=(-g_{\mu\nu}+\frac{Q_\mu Q_\nu}{t})/(t-M_{K^*}^{2}),
\end{eqnarray}
where $t\equiv Q^2$. The Feynman diagram is shown in Fig. \ref{abc}.

The $t$-channel amplitude in the quark model is given by
\begin{eqnarray}
\mathcal{M}^V_t=\mathcal{O}^t_V\frac{1}{t-M_{K^*}^{2}}e^{-(\mathbf{q}-\mathbf{k})^2/6\alpha^2},\label{t-v}
\end{eqnarray}
where $e^{-(\mathbf{q}-\mathbf{k})^2/6\alpha^2}$ is a quark model
form factor , $M_{K^*}$ is the mass vector meson $K^*$, and the
amplitude $\mathcal{O}^t_V$ is given by
\begin{eqnarray}
\mathcal{O}^t_V=G_v a[g^s_{t}(\mathcal{H}_0+\mathcal{H}_1
\mathbf{q}\cdot \mathbf{k})+g^v_t\mathcal{H}_2
i\vsig\cdot(\mathbf{q}\times \mathbf{k})]+\mathrm{tensor\ term}
,\label{tt-v}
\end{eqnarray}
in Eq. (\ref{tt-v}) we have defined
\begin{eqnarray}
\mathcal{H}_0&\equiv&E_0-\left[E_0
K_{is}+\left(\mathcal{K}_i+\frac{1}{6\mu_q}\right)(1+\mathcal{D})\right]k^2+\left[E_0\mathcal{K}_{fq}-\left(\mathcal{K}_f-\frac{1}{6\mu_q}\right)(1-\mathcal{T})\right]q^2,\\
\mathcal{H}_1&\equiv&E_0[\mathcal{K}_i\mathcal{K}_f-(\mathcal{K}_{fq}-\mathcal{K}_{is})]-(\mathcal{K}_i+\mathcal{K}_f)-(\mathcal{K}_i-\mathcal{K}_f)\mathcal{T}-\frac{1}{3\mu_q}\mathcal{T},\\
\mathcal{H}_2&\equiv&E_0[\mathcal{K}_f\mathcal{K}_i-(\mathcal{K}_{fq}-\mathcal{K}_{is})]-(\mathcal{K}_i+\mathcal{K}_f)-(\mathcal{K}_i
-\mathcal{K}_f)\mathcal{T}+\frac{1}{3}\left(\frac{1}{m_q}-\frac{1}{m_s}\right)\mathcal{T},
\end{eqnarray}
with
\begin{eqnarray}
\mathcal{K}_{fq}&=&\frac{1}{6m_q}\mathcal{K}_f,\
\mathcal{K}_{is}=\frac{1}{6m_s}\mathcal{K}_i,\\
\mathcal{T}&=&\frac{m^2_{\pi}-m^2_K}{t},\\
E_0&=&-(\omega_K+\omega_\pi)+(\omega_\pi-\omega_K)\mathcal{T}.
\end{eqnarray}
The $K^*$ exchange couplings, i.e. vector and tensor, can in
principle be determined by $K^*$ meson photoproduction. However, it
shows that the present experimental results from JLab and ELSA favor
quite differently the tensor coupling values. In $K^- p\to
\Sigma^0\pi^0$, the $K^*$ exchange is not a predominant transition
mechanism. We hence only consider the $t$-channel vector exchange,
but neglect the tensor term for simplicity.

In the Eq.(\ref{t-v}), we have defined $g^s_t\equiv \langle
N_f|\sum^3_{j=1}I^{K^-}_j|N_i\rangle$, and $g^v_t\equiv \langle
N_f|\sum^3_{j=1}\sigma_j I^{K^-}_j|N_i\rangle$, which can be deduced
from the quark model. Their values are listed in Tab. \ref{factor}.

Similarly, for the scalar meson $\kappa$-exchange,  the  $t$-channel
amplitude in the quark model is written as
\begin{eqnarray}
\mathcal{M}^S_t=\mathcal{O}^t_S\frac{1}{t-m^2_{\kappa}
}e^{-(\mathbf{q}-\mathbf{k})^2/6\alpha^2},
\end{eqnarray}
where $m_{\kappa}$ is the $\kappa$-meson mass, and $\mathcal{O}^t_S$
is given by
\begin{eqnarray}
\mathcal{O}^t_S\simeq\frac{g_{\kappa K \pi} g_{\kappa
qq}}{2m_\pi}(\omega_K\omega_\pi-\mathbf{q}\cdot
\mathbf{k})[g^s_{t}(\mathcal{A}_0+\mathcal{A}_1\mathbf{q}\cdot
\mathbf{k})+g^v_t \mathcal{A}_1 i\vsig\cdot(\mathbf{q}\times
\mathbf{k}) ],\label{t-s}
\end{eqnarray}
with
\begin{eqnarray}
\mathcal{A}_0\equiv1+\frac{1}{2m_q}\mathcal{K}_f\mathbf{q}^2-\frac{1}{2m_s}\mathcal{K}_i\mathbf{k}^2,\\
\mathcal{A}_1\equiv
\mathcal{K}_i\mathcal{K}_f-\frac{1}{2m_q}\mathcal{K}_f+\frac{1}{2m_s}\mathcal{K}_i.
\end{eqnarray}
In Eq.(\ref{t-s}), we have neglected the higher order terms.

\section{SEPARATION OF THE SINGLE RESONANCE CONTRIBUTIONS}\label{ss}

Note that, so far, we have separated out the amplitudes in terms of
the harmonic oscillator principle quantum number $n$, which are the
sum of a set of $SU(6)$ multiplets with the same $n$. To see the
contributions of individual resonances, we need to further separate
out the single-resonance-excitation amplitudes within each $n$ in
the $s$-channel. Since the resonances in the $u$-channel contribute
virtually and are generally suppressed by the kinematics, we treat
them as degenerate to $n$.

Function $F_s(n)$ in Eq.~(\ref{s-prop}) can be related to the
$s$-channel propagator in the infinitely-narrow-width limit:
 \begin{eqnarray}
 F_s(n)&=&\frac{2M_n}{s-(M_i^2+M_K^2+2nM_i\omega_h)}\equiv \frac{2M_n}{s-M_n^2} \ ,
\end{eqnarray}
where it has been assumed that $M_n^2\equiv
M_i^2+M_K^2+2nM_i\omega_h$, which is not a bad assumption for the
masses of an excited $n$-shell state.  $M_i$ denotes the initial
baryon mass.

Taking into account the width effects of the resonances, the
resonance transition amplitudes of the $s$-channel can be generally
expressed as \cite{qk3,Zhong:2007fx}
\begin{eqnarray}
\mathcal{M}^s_R=\frac{2M_R}{s-M^2_R+iM_R
\Gamma_R}\mathcal{O}_Re^{-(\textbf{k}^2+\textbf{q}^2)/6\alpha^2},
\label{stt}
\end{eqnarray}
and the $u$-channel as
\begin{eqnarray}
\mathcal{M}^u_n=-\frac{2M_n}{u-M^2_n}\mathcal{O}_n
e^{-(\textbf{k}^2+\textbf{q}^2)/6\alpha^2}.\label{utt}
\end{eqnarray}
In Eqs. (\ref{stt}) and (\ref{utt}), $\mathcal{O}_R$ is the
separated operators for individual resonances in the $s$-channel,
while $\mathcal{O}_n$ is the operator for a set of degenerate states
with the same $n$. In the $s$-channel of $K^-p\rightarrow
\Sigma^0\pi^0$, only the $\Lambda$ resonances are involved. Our
effort in the following subsections is to extract $\mathcal{O}_R$
for each $s$-channel resonance with $n<2$.

\subsection{$n=0$ shell resonances}

With $n=0$ the $\Lambda$-hyperon is the only state contributing to
the $s$-channel, and the amplitude can be written as
\begin{eqnarray}
\mathcal{M}_\Lambda^{s}&=&\mathcal{O}_\Lambda
\frac{2M_\Lambda}{s-M^2_\Lambda}e^{-(\textbf{k}^2+\textbf{q}^2)/6\alpha^2},
\end{eqnarray}
with
\begin{eqnarray}
\mathcal{O}_\Lambda=g_{s2}\textbf{A}_{out}\cdot\textbf{A}_{in}
+g_{v2}i\vsig
\cdot(\textbf{A}_{out}\times\textbf{A}_{in}),\nonumber\\
\end{eqnarray}
where $M_\Lambda$ is the $\Lambda$-hyperon mass.

\subsection{$n=1$ shell resonances}

Both $S$- and $D$-wave resonances contribute to the $s$-channel
amplitude with $n=1$. Note that the spin-independent amplitude for
$D$-waves is proportional to the Legendre function
$P^0_2(\cos\theta)$, and the spin-dependent amplitude is in
proportion to $\frac{\partial}{\partial \theta}P^0_2(\cos\theta)$.
Moreover, the $S$-wave amplitude is independent of the scattering
angle. Thus, the $S$- and $D$-wave amplitudes can be separated out
easily as follows,
\begin{eqnarray}
\mathcal{O}_S&=&-\frac{1}{2}g_{s2}\left(|\mathbf{A}_{out}|\cdot|\mathbf{A}_{in}|\frac{|\mathbf{k}||\mathbf{q}|}{9\alpha^2}
-\frac{\omega_K}{6\mu_q}\mathbf{A}_{out}\cdot
\mathbf{q}-\frac{\omega_\pi}{3m_q}\mathbf{A}_{in}\cdot
\mathbf{k}+\frac{\omega_\pi\omega_K}{2m_q\mu_q}\frac{\alpha^2}{3}\right),\\
\mathcal{O}_D&=&-\frac{1}{2}g_{s2}|\mathbf{A}_{out}|\cdot|\mathbf{A}_{in}|(3\cos^2\theta-1)
\frac{|\mathbf{k}||\mathbf{q}|}{9\alpha^2}-\frac{1}{2}g_{v2}i\vsig
\cdot(\textbf{A}_{out}\times\textbf{A}_{in})\frac{\mathbf{k}\cdot\mathbf{q}}{3\alpha^2}.
\end{eqnarray}

In the NRCQM, the $n=1$ shell contains three different
representations, i.e. $[\textbf{70},^2\textbf{1}]$,
$[\textbf{70},^2\textbf{8}]$ and $[\textbf{70},^4\textbf{8}]$. The
two low-lying $\Lambda$-resonances, $\Lambda(1405)S_{01}$ and
$\Lambda(1520)D_{03}$, are classified to be flavor singlet states of
$[\textbf{70},^2\textbf{1}]$, and they have no counterparts in the
nucleon spectrum. The $\Lambda(1670)S_{01}$ and
$\Lambda(1690)D_{03}$ are interpreted as multiplets of
$[\textbf{70},^2\textbf{8}]$, which are octet partners of the
nucleon resonances $S_{11}(1535)$ and $D_{13}(1520)$. Usually the
$\Lambda(1800)S_{01}$ and $\Lambda(1830)D_{05}$ are classified as
multiplets of $[\textbf{70},^4\textbf{8}]$, among which the $D_{03}$
state has not yet been found in experiment. In the $SU(6)\otimes
O(3)$ quark model, the contributions of $[\textbf{70},^4\textbf{8}]$
are forbidden in $K^-p\rightarrow \Sigma^0 \pi^0$ due to the
so-called ``$\Lambda$-selection rule"
\cite{Zhao:2006an,Isgur:1978xb,Hey:1974nc}. Thus, for the $S$-wave,
only the resonances, $\Lambda(1405)S_{01}$ and
$\Lambda(1670)S_{01}$, contribute to the reactions, and for the
$D$-waves, $\Lambda(1520)D_{03}$ and $\Lambda(1690)D_{03}$.

The separated amplitudes for the $S$- and $D$-wave can thus be
re-written as
\begin{eqnarray}
\mathcal{O}_S&=&[g_{S_{01}(1405)}+g_{S_{01}(1670)}]\mathcal{O}_S,\\
\mathcal{O}_D&=&[g_{D_{03}(1520)}+g_{D_{03}(1690)}]\mathcal{O}_D,
\end{eqnarray}
where the factor $g_R$ ($R=S_{01}(1405)$, etc) represents the
resonance transition strengths in the spin-flavor space, and is
determined by the matrix element $\langle
N_f|H_\pi|N_j\rangle\langle N_j|H_K|N_i\rangle$. Their relative
strengths can be explicitly determined by the following relations
\begin{eqnarray}
\frac{g_{S_{01}(1405)}}{g_{S_{01}(1670)}}=\frac{\langle
N_f|I^{\pi}_3\vsig_3|S_{01}(1405)\rangle\langle
S_{01}(1405)|I^K_3\vsig_3|N_i\rangle}{\langle
N_f|I^{\pi}\vsig_3|S_{01}(1670)\rangle\langle
S_{01}(1670)|I^K_3\vsig_3|N_i\rangle},\label{abcd}\\
\frac{g_{D_{03}(1520)}}{g_{D_{03}(1690)}}=\frac{\langle
N_f|I^{\pi}_3\vsig_3|D_{03}(1520)\rangle\langle
D_{03}(1520)|I^K_3\vsig_3|N_i\rangle}{\langle
N_f|I^{\pi}\vsig_3|D_{03}(1690)\rangle\langle
D_{03}(1690)|I^K_3\vsig_3|N_i\rangle}.
\end{eqnarray}
On the condition of no configuration mixing among these states, we
have
$g_{S_{01}(1405)}/g_{S_{01}(1670)}=g_{D_{03}(1520)}/g_{D_{03}(1690)}=-3$.
However, the admixtures of different configurations usually occur in
physical states with the same quantum number due to spin-dependent
forces~\cite{Isgur78,Loring:2001ky}. We shall see in Sec.\ref{cy}
that configuration mixing may exist between the $S$-wave
$S_{01}(1405)$ and $S_{01}(1670)$ in this reaction. By allowing the
data to constraint the relative partial strengths, i.e.
$g_{S_{01}(1405)}/g_{S_{01}(1670)}$, we can extract the mixing angle
as a leading order result.

With the same method, we can separate the amplitudes in $n=2$ shell
as well, the detail can be found in our previous work
\cite{Zhong:2007fx}. In this work, the higher resonances (i.e.
$n\geq 2$) are treated as degenerate since they are less important
in the beam momentum region $P_K\lesssim 800$ MeV/c where high
precision data are available.

\begin{table}[ht]
\caption{Various g and $g_R$ factors defined in this work and
extracted in the symmetric quark model. } \label{factor}
\begin{tabular}{|c|c|c|c|c|c|c|c|c }\hline\hline
factor & value               &    factor  &  value         \\
\hline
$g^u_{s1}$ & 1/2                &  $g^s_{t}$   & $\sqrt{2}/2 $          \\
$g^u_{s2}$ & 2/3                &  $g^v_{t}$   &  $-\sqrt{2}/6$         \\
$g^u_{v1}$ &  -1/6              &  $g_{S_{01}(1405)}$   &  3/2          \\
$g^u_{v2}$ & -1                 &  $g_{S_{01}(1670)}$ & -1/2  \\
$g_{s2}$   & 2/3                &  $g_{D_{03}(1520)}$ &  3/2    \\
$g_{v2}$   &  1                 &  $g_{D_{03}(1690)}$  & -1/2\\
\hline
\end{tabular}
\end{table}

\section{calculation and analysis} \label{cy}

\subsection{Parameters}

With the transition amplitudes derived from the previous section,
the differential cross section can be calculated,
\begin{eqnarray}
\frac{d\vsig}{d\Omega}=\frac{(E_i+M_i)(E_f+M_f)}{64\pi^2
s}\frac{|\textbf{q}|}{|\textbf{k}|}\frac{1}{2}
\sum_{\lambda_i,\lambda_f}\left|\left[\frac{\delta^2}{f_\pi
f_K}(\mathcal{M}_s+\mathcal{M}_u)+\mathcal{M}^V_t+\mathcal{M}^S_t\right]_{\lambda_f,\lambda_i}\right|^2
,
\end{eqnarray}
where $\lambda_i=\pm 1/2$ and $\lambda_f=\pm 1/2$ are the helicities
of the initial and final state baryons, respectively; $\delta$ is a
global parameter accounting for the flavor symmetry breaking effects
arising from the $quark-meson$ couplings, and will be determined by
experimental data; $f_\pi$ and $f_K$ are the $\pi$- and $K$-mesons
decay constants, respectively.

To take into account the relativistic effects, we introduce Lorentz
boost factors in the spatial part of the amplitudes as done in Refs.
\cite{qkk,Zhong:2007fx}, i.e.
\begin{eqnarray}
\mathcal{O}_i(\textbf{k},\textbf{q})\rightarrow \gamma_k \gamma_q
\mathcal{O}_i(\textbf{k}\gamma_k, \textbf{q} \gamma_q),
\end{eqnarray}
where $\gamma_k=M_i/E_i$ and $\gamma_q=M_f/E_f$.

We also introduce an energy-dependent width for the resonances in
order to take into account the off-mass-shell effects in the
reaction~\cite{qkk,qk3,qk4}:
\begin{eqnarray}
\Gamma(\textbf{q})=\Gamma_R\frac{\sqrt{s}}{M_R}\sum_i x_i
\left(\frac{|\textbf{q}_i|}{|\textbf{q}^R_i|}\right)^{2l+1}
\frac{D(\textbf{q}_i)}{D(\textbf{q}^R_i)},
\end{eqnarray}
where $|\textbf{q}^R_i|=((M_R^2-M_b^2+m_i^2)/4M_R^2-m_i^2)^{1/2}$,
and $|\textbf{q}_i|=((s-M_b^2+m_i^2)/4s-m_i^2)^{1/2}$; $x_i$ is the
branching ratio of the resonance decaying into a meson with mass
$m_i$ and a baryon with mass $M_b$, and $\Gamma_R$ is the total
decay width of the $s$-channel resonance with mass $M_R$.
$D(\textbf{q})=e^{-\textbf{q}^2/3\alpha^2}$ is a fission barrier
function.

In the calculation, the universal value of harmonic oscillator
parameter $\alpha=0.4$ GeV is adopted. The masses of the $u$, $d$,
and $s$ constituent quarks are set as $m_u=m_d=330$ MeV, and
$m_s=450$ MeV, respectively. The decay constants for $\pi$, and $K$
are $f_\pi=132$ MeV and $f_K=160$ MeV, respectively.


Coupling constants in the $t$-channel transitions, i.e. $G_v$, $a$,
$g_{\kappa K\pi}$ and $g_{\kappa qq}$, can be determined by other
experimental data. For instance, $G_v$ can be determined by $K^*\to
K\pi$~\cite{PDG}, while vector coupling $a$ can be extracted from
$K^*$ photoproduction~\cite{JLab-Kstar,elsa-kstar,zhao-kstar}. As
shown by Refs.~\cite{JLab-Kstar,elsa-kstar}, coupling $a$ has a
value of about 3, but with quite significant uncertainties. As $G_v$
and $a$ appear simultaneously in the product of $G_v a$, we find
that $G_va=38$ is a reasonable value for the $K^*$ exchange. Note
that within the uncertainties of $K^*N\Sigma$ coupling, this value
can be regarded as reasonable. The value of $g_{\kappa K\pi }$
predicted by QCD sum rules is $g_{\kappa K\pi }\simeq 4$, which is
compatible with the value extracted from the data
\cite{Brito:2004tv}. This implies that the $\kappa qq$ coupling
constant is $g_{\kappa qq}\simeq 5$, which also turns to be
reasonable.

Parameters in the $s$- and $u$-channel will be determined by fitting
the cross section data.  So far, there are 63 datum points of
differential cross section at seven momentum beams between 514 and
687 MeV/c available~\cite{Manweiler:2008zz}. By fitting this datum
set, we find $\delta\simeq 1.55$ accounting for flavor symmetry
breaking effects, and resonance parameters are also determined and
listed in the Tab. \ref{parameter}. From the table, we see that all
the resonance parameters roughly agree with the PDG values. The
preferred Breit-Wigner mass of the $\Lambda(1405)S_{01}$ is $1420$
MeV, which is about 10 MeV larger than the upper limit of the PDG
suggestion~\cite{PDG}. To fit the total cross section, we find the
widths of $\Lambda(1520)D_{03}$ should have a narrower width
$\Gamma\simeq 8$ MeV, which is only half of the PDG value. The
fitted  mass and width for $\Lambda(1670)S_{01}$ are $M=1697$ and
$\Gamma=65$ MeV, respectively, which are also slightly larger than
the PDG suggestions. For the $n=2$ shell we take a degenerate mass
and width as $M=1850$ MeV and $\Gamma=100$ MeV since in the low
energy region contributions from the $n=2$ shell are not
significant.

\begin{table}[ht]
\caption{Breit-Wigner masses $M_R$ (in MeV) and widths $\Gamma_R$
(in MeV) for the resonances in the $s$-channel. States in the $n=2$
shell are treated as degenerate to $n$.
} \label{parameter}
\begin{tabular}{|c|c|c||c|c|c| }\hline\hline
resonance &\ \  $M_R$ \ \ & \ \ $\Gamma_R$ \ \ &\ \ $M_R$ (PDG)\ \ & \ \ $\Gamma_R$ (PDG)\ \  \\
\hline
$S_{01}(1405)$& 1420    &48  & $1406\pm 4 $   &$50\pm 2$   \\
$S_{01}(1670)$& 1697    &65  &  $1670\pm 10$    & $25\sim50$  \\
$D_{03}(1520)$& 1520    &8   &  $1520\pm 1$   &$16\pm 1$ \\
$D_{03}(1690)$& 1685    &63  &  $1690\pm 5$    &$60\pm 10$ \\
n=2           & 1850    &100 &      &   \\
\hline
\end{tabular}
\end{table}

In the $u$-channel, the intermediate states are the nucleon and its
resonances. We find that contributions from the $n\geq 1$ shell are
negligibly small, and are insensitive to the degenerate masses and
widths for these shells. In this work, we take $M_1=1650$ MeV
($M_2=1750$ MeV), $\Gamma_1=230$ MeV ($\Gamma_2=300$ MeV) for the
degenerate mass and width of $n=1$ ($n=2$) shell nucleon resonances,
respectively.

The last parameter we consider is the relative strength
$g_{S_{01}(1405)}/g_{S_{01}(1670)}$. The data favor a much larger
value for $g_{S_{01}(1405)}$ relative to $g_{S_{01}(1670)}$. In
another word, a much stronger $S$-wave contribution is needed in the
explanation of the experimental data. We thus empirically adjust the
relative strength between $S_{01}(1405)$ and $S_{01}(1670)$ by a
mixing angle (see Sec.\ref{cm}). This could be evidence that the
single quark interaction picture fails in the description of the
dominant $S$-wave amplitude.

\subsection{Configuration mixing}\label{cm}

\begin{center}
\begin{figure}[ht]
\centering \epsfxsize=10 cm \epsfbox{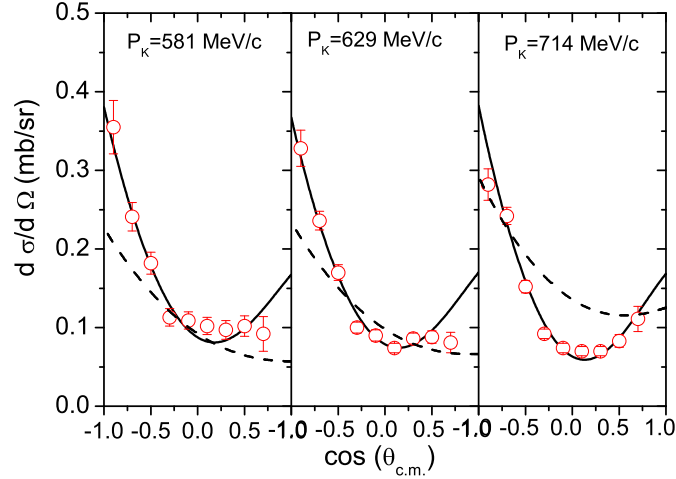} \caption{(Color
online) Comparisons of the differential cross sections between with
(solid curves) and without configuration mixings (dashed curves) for
the $\Lambda(1405)$ and $\Lambda(1670)$, respectively. }\label{cmf}
\end{figure}
\end{center}

In the calculations, we find that the relative strength
$g_{S_{01}(1405)}/g_{S_{01}(1670)}$ is crucial for reproducing the
angular distributions in the differential cross sections. With no
configuration mixing, i.e. $g_{S_{01}(1405)}/g_{S_{01}(1670)}=-3$,
the data can not be well explained as shown by the dashed curves in
Fig. \ref{cmf}.

As we know, the configuration mixing will bring uncertainties to
this value, thus we determine it by fitting the data. When we take
$g_{S_{01}(1405)}/g_{S_{01}(1670)}\simeq -9$, the data can be
reasonably reproduced (see the solid curves in Fig. \ref{cmf}),
which indicates that the configuration mixing in $S_{01}(1405)$ and
$ S_{01}(1670)$ is needed.  If we take the
$g_{D_{03}(1520)}/g_{D_{03}(1690)}$ as a free parameter, the fitted
value do not change obviously compared with the value of no
configuration mixing. Thus, in the calculations, we do not
considered the configuration mixing in $D_{03}(1520)$ and
$D_{03}(1690)$.

We empirically introduce a mixing angle between
$[\textbf{70},^2\textbf{1}]$ and $[\textbf{70},^2\textbf{8}]$ within
the physical states $S_{01}(1405)$ and $S_{01}(1670)$, i.e.
\begin{eqnarray}
|S_{01}(1405)\rangle=\cos(\theta)|\textbf{70},^2\textbf{1}
\rangle-\sin(\theta)|\textbf{70},^2\textbf{8} \rangle,\\
|S_{01}(1670)\rangle=\sin(\theta)|\textbf{70},^2\textbf{1}
\rangle+\cos(\theta)|\textbf{70},^2\textbf{8} \rangle.
\end{eqnarray}
Inserting these wave functions into Eq.(\ref{abcd}), we have
\begin{eqnarray}\label{mix-coup}
\frac{g_{S_{01}(1405)}}{
g_{S_{01}(1670)}}=\frac{[3\cos(\theta)-\sin(\theta)][\cos(\theta)+\sin(\theta)]}
{[3\sin(\theta)+\cos(\theta)][\sin(\theta)-\cos(\theta)]}\label{rt},
\end{eqnarray}
which is a function of the mixing angle $\theta$.

In order to study the relation between the relative coupling
strength $g_{S_{01}(1405)}/g_{S_{01}(1670)}$ and mixing angle
$\theta$, we define a function of $\theta$ as
\begin{eqnarray}
f(\theta)=[3\cos(\theta)-\sin(\theta)][\cos(\theta)+\sin(\theta)]
-\frac{g_{S_{01}(1405)}}{
g_{S_{01}(1670)}}[3\sin(\theta)+\cos(\theta)][\sin(\theta)-\cos(\theta)].
\end{eqnarray}
For a given ratio $g_{S_{01}(1405)}/g_{S_{01}(1670)}$, the mixing
angle $\theta$ can be determined at $f(\theta)=0$. One can easily
check that the ratio $g_{S_{01}(1405)}/g_{S_{01}(1670)}=-3$ leads to
$\theta=0^\circ$, i.e. no configuration mixing between
$[\textbf{70},^2\textbf{1}]$ and $[\textbf{70},^2\textbf{8}]$.

\begin{center}
\begin{figure}[ht]
\centering \epsfxsize=8 cm \epsfbox{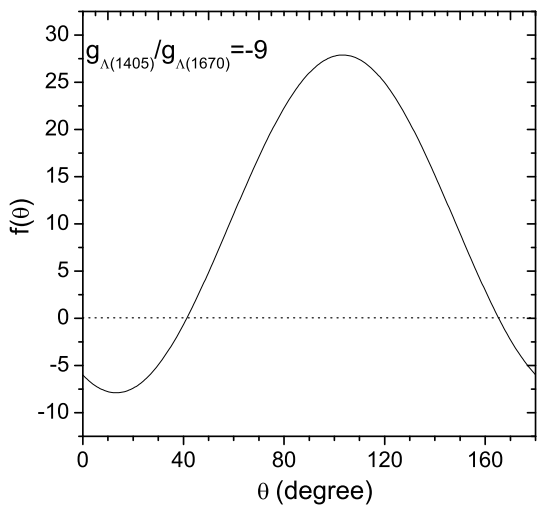} \caption{ The
evolution of function $f(\theta)$ in terms of the mixing angle
$\theta$ is shown. The values of $\theta$ corresponding to
$f(\theta)=0$ are the mixing angles for $g_{S_{01}(1405)}/
g_{S_{01}(1670)}=-9$, which are found to be $\theta\simeq 41^\circ$
and $\theta\simeq 165^\circ$.}\label{figq}
\end{figure}
\end{center}

With the fitted value $g_{S_{01}(1405)}/ g_{S_{01}(1670)}=-9$, the
$f(\theta)$ as a function of $\theta$ is shown in Fig. \ref{figq}.
The mixing angle can then be extracted at $f(\theta)=0$. From the
figure, we find that two mixing angles, $\theta\simeq41^\circ$ and
$165^\circ$, satisfy the condition $f(\theta)=0$ with
$g_{S_{01}(1405)}/ g_{S_{01}(1670)}=-9$.

With $\theta=41^\circ$, the admixtures of flavor singlet
$[\textbf{70},^2\textbf{1}]$ and flavor octet
$[\textbf{70},^2\textbf{8}]$ in the $\Lambda(1405)$ amount to $57\%$
and $43\%$, respectively. With $\theta=165^\circ$, the
$\Lambda(1405)$ is dominantly $[\textbf{70},^2\textbf{1}]$ with a
wave function density of $\sim 93\%$, while admixture of
$[\textbf{70},^2\textbf{1}]$ in $\Lambda(1670)$ is only $\sim 7\%$.
The recent relativistic quark model study suggests that for the
$\Lambda(1405)$ the admixtures of singlet
$[\textbf{70},^2\textbf{1}]$ and octet $[\textbf{70},^2\textbf{8}]$
are $\sim 70\%$ and $\sim 30\%$, respectively, and for the
$\Lambda(1670)$, the admixture of $[\textbf{70},^2\textbf{8}]$ is
$\sim 62\%$ and that of $[\textbf{70},^2\textbf{1}]$ is $\sim 26\%$
\cite{Loring:2001ky}, which is compatible with the results with
$\theta=41^\circ$. It is interesting to note that this feature that
the $\Lambda(1405)$ and $\Lambda(1670)$ as mixed states dominated by
the singlet and octet, respectively, is also obtained by the coupled
channel studies based on U$\chi$PT~\cite{Jido:2003cb}.

Furthermore, Eq. (\ref{mix-coup}) allows us to investigate the
ratios of the couplings to the $\bar{K}N$ and $\pi \Sigma$ channel
for the states $S_{01}(1405)$ and $S_{01}(1670)$ with the following
relations:
\begin{eqnarray}
\frac{g_{S_{01}(1405)\bar{K}N}}{
g_{S_{01}(1670)\bar{K}N}}=\frac{\cos(\theta)+\sin(\theta)}
{\sin(\theta)-\cos(\theta)},\\
\frac{g_{S_{01}(1405)\pi \Sigma}}{ g_{S_{01}(1670)\pi\Sigma}}=\frac{
3\cos(\theta)-\sin(\theta) } { 3\sin(\theta)+\cos(\theta) }.
\end{eqnarray}

If we take the mixing angle $\theta=41^\circ$ we have
\begin{eqnarray}
\left|\frac{g_{S_{01}(1405)\bar{K}N}}{
g_{S_{01}(1670)\bar{K}N}}\right|\simeq 14,\
\left|\frac{g_{S_{01}(1405)\pi \Sigma}}{
g_{S_{01}(1670)\pi\Sigma}}\right|\simeq 0.6.
\end{eqnarray}
This solution  is in agreement with the UChPT model prediction
\cite{Oset:2001cn}, which also prefers a much stronger coupling of
the $S_{01}(1405)$ to the $\bar{K} N$ channel than the
$S_{01}(1670)$.

On the other hand, if the mixing angle is taken as
$\theta=165^\circ$ it gives
\begin{eqnarray}
\left|\frac{g_{S_{01}(1405)\bar{K}N}}{
g_{S_{01}(1670)\bar{K}N}}\right|\simeq 0.58,\
\left|\frac{g_{S_{01}(1405)\pi \Sigma}}{
g_{S_{01}(1670)\pi\Sigma}}\right|\simeq 17.
\end{eqnarray}
In order to determine the mixing angle and the couplings for these
two $S_{01}$ states, a coherent study of the photoproduction $\gamma
p\to K^+ \Lambda(1405)$ and $\gamma p\to K^+ \Lambda(1670)$ would be
needed.

\subsection{Differential cross section}\label{dc}

\begin{center}
\begin{figure}[ht]
\centering \epsfxsize=12 cm \epsfbox{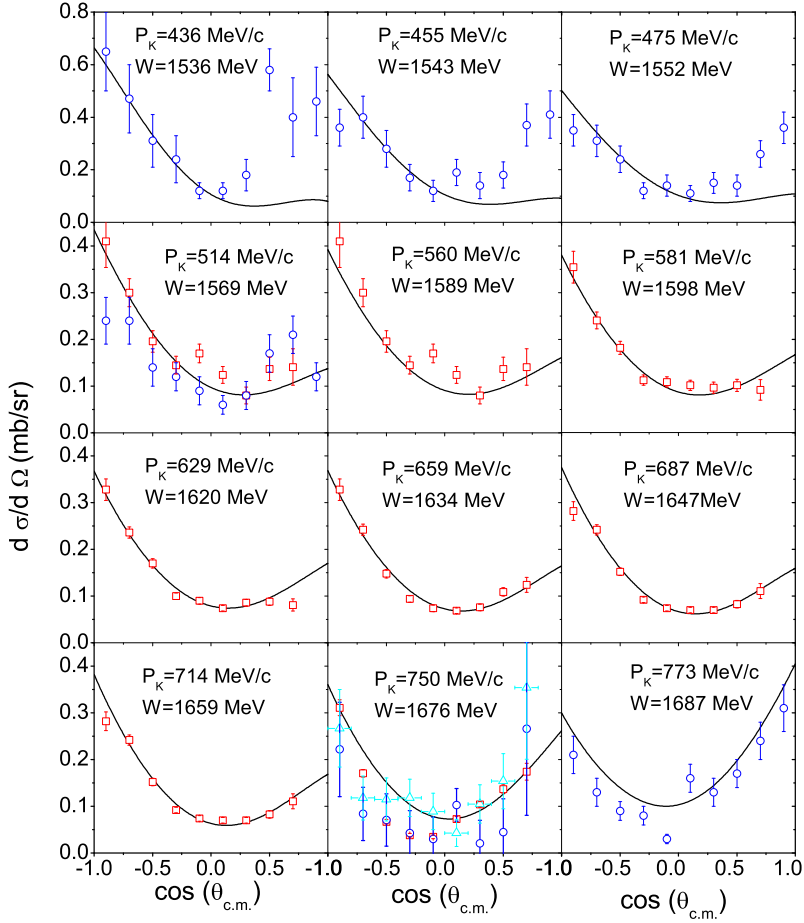} \caption{ (Color
online) Differential cross sections for $P_K=475\sim 775$ MeV/c
(i.e. $W=1536\sim 1687$ MeV). Data are from \cite{Manweiler:2008zz}
(open squares), \cite{Baxter:1974zs} (open up-triangles) and
\cite{armen:1970zh} (open circles). }\label{fig-1}
\end{figure}
\end{center}

In Fig. \ref{fig-1}, the differential cross sections are shown at
different center mass energies (beam momenta) from $W=1536$ MeV
($P_K=436$ MeV/c) to $W=1687$ MeV ($P_K=773$ MeV/c). The
experimental
data~\cite{Manweiler:2008zz,armen:1970zh,London:1975av,Baxter:1974zs,Berley:1996zh,Mast:1974sx}
are also included for a comparison. As shown by the solid curves,
the overall agreement with the experimental data is rather good.
However, we also note that the theoretical results seem to slightly
underestimate the differential cross sections at forward angles at
$W=1536\sim 1552$ MeV, which is just around the
$\Lambda(1520)D_{03}$ production threshold. Notice that the
experimental data possess quite large uncertainties, improved
measurement in this energy region is needed to clarify the
discrepancies.

\begin{center}
\begin{figure}[ht]
\centering \epsfxsize=9 cm \epsfbox{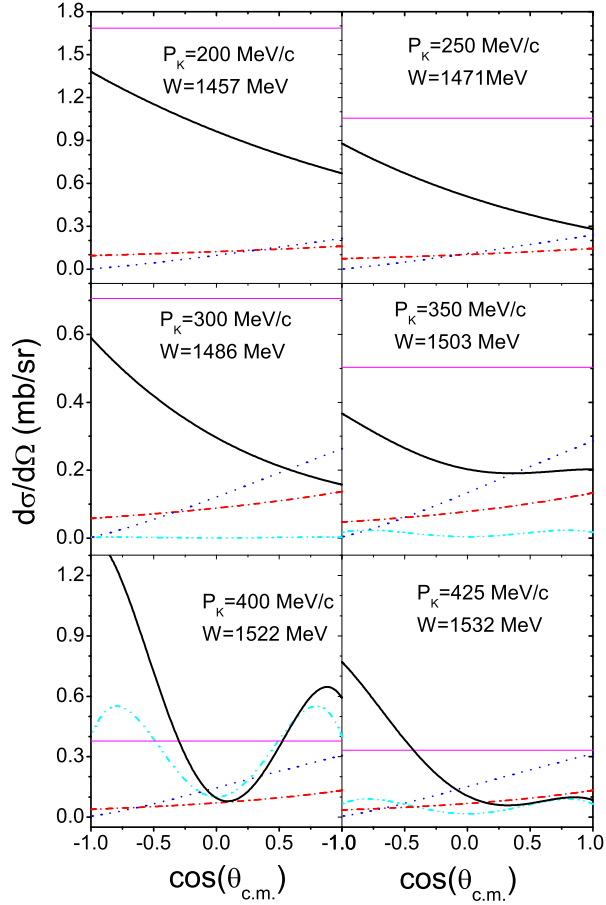} \caption{(Color
online) Differential cross sections at six energies in a range of
$P_K=200\sim 425$ MeV/c (i.e. $W=1457\sim 1532$ MeV). The bold solid
curves are given by the full model calculations. The thin lines,
dashed, dash-dotted and dash-dot-dotted curves stand for the
exclusive cross sections for the $S_{01}(1405)$, $u$-channel,
$t$-channel $K^*$-exchange, and the $D_{03}(1520)$, respectively.
}\label{fig-4}
\end{figure}
\end{center}

\begin{center}
\begin{figure}[ht]
\centering \epsfxsize=9 cm \epsfbox{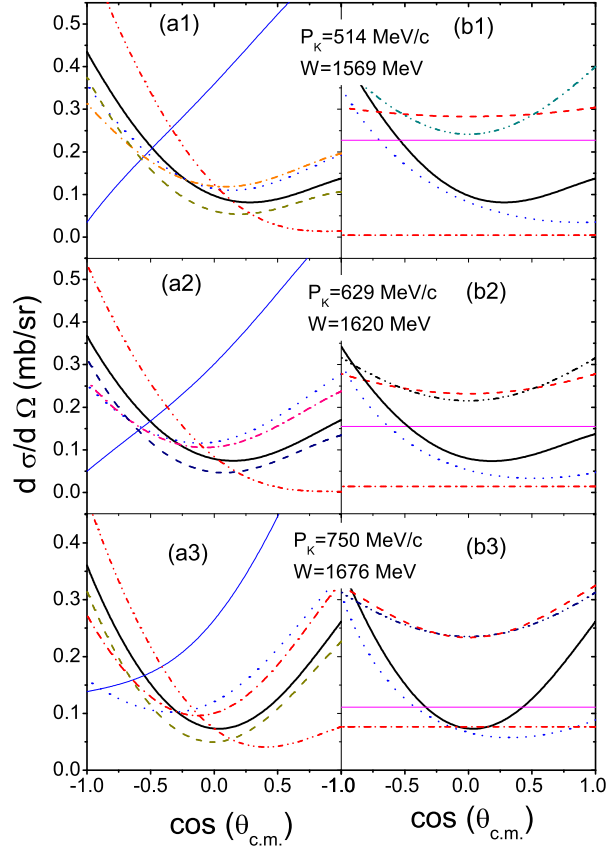} \caption{ (Color
online) Cross sections of exclusive channels or individual
resonances are shown at $P_K=514$, 629 and 750 MeV/c, respectively.
The bold solid curves are for the full model calculations. In the
left panel, i.e. (a1)-(a3), the dashed, dash-dotted, and
dash-dot-dotted, dotted and thin solid curves are for the results
given by switching off the contributions from the $\Lambda$ pole,
$t$-channel $\kappa$-exchange, $t$-channel $K^*$-exchange,
$\Lambda(1670)S_{01}$, and $\Lambda(1405)S_{01}$, respectively. In
the right panel, i.e. (b1)-(b3), the dashed, dash-dot-dotted and
dotted curves correspond to the interferences of the
$\Lambda(1690)D_{03}$, $\Lambda(1520)D_{03}$ and $u$-channel with
the $S$-wave amplitudes, respectively. The thin solid lines and the
dash-dotted lines stand for the exclusive cross sections of the
$\Lambda(1405)S_{01}$ and $\Lambda(1670)S_{01}$,
respectively.}\label{fig-2}
\end{figure}
\end{center}

To the low energy region, i.e. $W=1457\sim 1532$ MeV (or
$P_K=200\sim 425$ MeV/c), there are no data for the differential
cross sections available from experiment. This is the region that
the low-lying $\Lambda(1405)S_{01}$ dominates. Therefore, we plot in
Fig. \ref{fig-4} the cross sections given by our model in
association with exclusive cross sections by single resonance
excitations or transitions. We also carry out such a decomposition
for the differential cross sections in the region of $W=1569 \sim
1676$ MeV in Fig. \ref{fig-2}.

In Figs. \ref{fig-4}, the solid curves are the full calculations of
the model. The thin horizontal lines denote the contributions from
the $\Lambda(1405)S_{01}$. Interestingly, the $\Lambda(1405)S_{01}$
appears to be predominant and even larger than the full results. It
implies that large cancelations exist between the
$\Lambda(1405)S_{01}$ amplitude and other transitions.

The dotted curves in Fig.  \ref{fig-4} are contributions from the
$u$-channel transition. It presents an enhancement at forward angles
though the $u$-channel propagator will generally suppress the
forward-angle cross sections. Reason for this enhancement is due to
the cancelations occur within the term of ${\bf B}_{in}\cdot {\bf
B}_{out}$ at backward angles. Meanwhile, the $u$-channel will
provide an important destructive interference with the
$\Lambda(1405)S_{01}$, and lower the differential cross sections at
the forward direction.

The dash-dotted curves in Fig.~\ref{fig-4} represent contributions
from the $t$-channel $K^*$ exchange, which are also forward-angle
enhanced. This contribution deceases with the energies and provides
an essentially important interference in the amplitudes. As shown in
Fig. \ref{fig-2}(a) by the dash-dot-dotted curves, its interferences
with the rest mechanisms will enhance the forward-angle cross
sections, but suppress the backward ones. In contrast, the overall
effects from the $t$-channel $\kappa$ exchange are rather small.

At $W=1522$ MeV (i.e. $P_K \sim 400$ MeV/c), the contributions from
the on-shell $D_{03}(1520)$ can be seen clearly by its interference
which significantly changes the shape of the differential cross
section. However, in the energies away from its mass, the $D$-wave
effects die out quickly.

Further study of the individual transitions are presented in Fig.
\ref{fig-2}, where in the left panel the cross sections are given by
removing one of the transition amplitudes from contributing, while
in the right panel cross sections are given by single transitions
interfering with the $S$-waves, i.e. $\Lambda(1405)S_{01}$ and
$\Lambda(1670)S_{01}$. First in the right panel, the two horizontal
lines, thin solid and dash-dotted, are exclusive cross sections for
the $\Lambda(1405)S_{01}$ and $\Lambda(1670)S_{01}$, respectively.
In the energy region of $W=1569\sim 1676$ MeV, the
$\Lambda(1405)S_{01}$ is no longer a dominant amplitude though its
contribution is still significant. By adding the
$\Lambda(1690)D_{03}$, $\Lambda(1520)D_{03}$, and the $u$-channel to
the $S$-waves, their effects are shown by the dashed,
dash-dot-dotted, and dotted curves, respectively. It is interesting
to see the role played by the $u$-channel, of which the interference
contributes to the creation of the backward enhancement.

On the left panel, the thin lines shows the effects without the
$\Lambda(1405)S_{01}$, which are strongly forward peaking.
Alternatively, this shows how important the $\Lambda(1405)S_{01}$ is
in this reaction. The other drastic effects are illustrated by the
dash-dot-dotted curves, which are generated by removing the
$t$-channel $K^*$ exchange. As discussed earlier, it contributes to
the forward enhancement and suppresses the backward cross sections.
As shown by the dashed, dotted, and dashed-dotted curves,
interfering effects from $\Lambda$ pole, $\Lambda(1670)S_{01}$, and
$t$-channel $\kappa$ can also be identified. In particular, it shows
that the $\kappa$ exchange interferes with the other amplitudes in
an opposite behavior in comparison with the $K^*$. It suppresses the
forward-angle cross sections but enhances the backward ones.

It is interesting to compare this study with $\pi^-p\rightarrow \eta
n$ \cite{Zhong:2007fx}, where the cross section is also dominated by
the $S$-wave near threshold, but the angular distribution is mainly
controlled by the $S$- and $D$-wave interferences. In
$K^-p\rightarrow \pi^0\Sigma^0$, we find that the interferences
between the $S$-wave and the $u$-channel are more crucial in the
energy region $P_K\gtrsim 520$ MeV/c. The $D$-wave interferences
become restricted to a relatively narrow energy region due to the
narrow width of $\Lambda$ states. It should also be recognized that
since only the amplitude ${\cal M}_2^s$ can contribute, the
$s$-channel interferences from the $\Lambda$ pole is not as
significant as the $u$-channel.

\subsection{Total cross section}

\begin{figure}[ht]
\centering \epsfxsize=8 cm \epsfbox{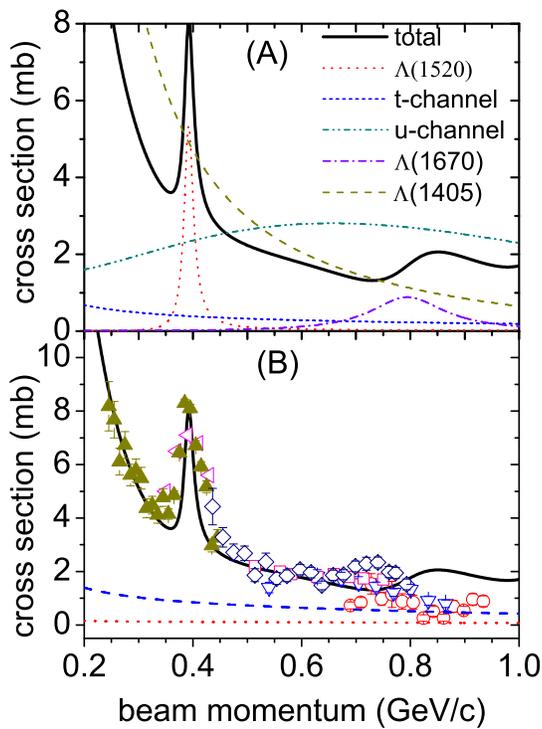} \caption{(Color
online) Total cross section as a function of the beam momentum
$P_K$. The solid curves are the full model calculations. Data are
from Refs. \cite{Baxter:1974zs} (open circles), \cite{Mast:1974sx} (
up-triangles), \cite{armen:1970zh} (open diamonds),
\cite{Berley:1996zh} (left-triangles), \cite{London:1975av}
(down-triangles), and \cite{Manweiler:2008zz} (squares). In (A),
exclusive cross sections for the $\Lambda(1405)S_{01}$,
$\Lambda(1670)S_{01}$, $\Lambda(1520)D_{03}$, $t$-channel, and
$u$-channel are indicated by different lines, respectively. In (B),
the dotted and dashed curves correspond to the exclusive cross
sections for the $t$-channel $\kappa$ and $K^*$-exchange,
respectively. }\label{fig-3}
\end{figure}

The total cross section as a function of the beam momentum is
plotted in Fig. \ref{fig-3} to compare with experimental
data~\cite{Baxter:1974zs,Mast:1974sx,armen:1970zh,Berley:1996zh,London:1975av,Manweiler:2008zz}.
To see the contributions of exclusive transitions, their cross
sections are also plotted. It shows that our theoretical
calculations agree well with the experimental data up to $P_K<800$
MeV.

Towards the low-energy limit, the total cross section exhibits a
steep enhancement which is due to the dominant
$\Lambda(1405)S_{01}$. The dashed curve shows the exclusive cross
section of the $\Lambda(1405)S_{01}$, which is larger than the total
cross section of the full calculations. The $u$-channel also turns
out to be a major contributor to the cross sections, and is a main
background in the whole momentum region. It becomes even larger than
the other transitions above $P_K>500$ MeV, and its interference with
the $S$-wave amplitudes governs the momentum-dependent behavior of
the cross section except for the resonance excitations by the
$\Lambda(1520)D_{03}$, which produces a sharp peak in the total
cross section. The importance of the $u$-channel contributions are
also stressed in the U$\chi$PT calculations
\cite{Oller:2006jw,Oller:2005ig}. It is found there that by
switching off the $I=1$ resonances,  the results change quite
significantly near threshold.

To reproduce this peak, it requires that the $\Lambda(1520)D_{03}$
has a narrow width $\Gamma\simeq 8$ MeV, which is about a factor 2
smaller than the PDG value. The contributions of the
$\Lambda(1670)S_{01}$ are also visible around $P_K=0.8\pm 1$ GeV/c.
When the beam momentum $P_K\gtrsim 800$ MeV, the model predictions
start to become worse, which indicates that the treatment of the
resonances of $n=2$ shell as degenerate is no longer applied, and
more realistic approach should be introduced. Because of the lack of
accurate data in this momentum region, we do not discuss the higher
resonances in the $n\geq 2$ shells in this work.

The $t$-channel $K^*$ and $\kappa$ exchanges are also shown, and
they both decrease with the increase of the beam momentum.
Furthermore, the $K^*$ exchange is much larger than the $\kappa$
exchange.

\section{summary and discussion}\label{sum}

In this work we have studied the reaction $K^-p\rightarrow
\Sigma^0\pi^0$ at low energies within a chiral quark model. With a
limited number of parameters, we can describe the differential cross
sections and cross sections which are in a good agreement with the
data. In the low energy region, i.e., $P_K< 800$ MeV/c, the $n=1$
shell resonances $\Lambda(1405)S_{01}$, $\Lambda(1520)D_{03}$ and
$\Lambda(1670)S_{01}$ are found to play important roles in the
reactions, and the $n\geq 2$ shell resonance contributions are
negligible small.

The $\Lambda(1405)S_{01}$ is very crucial in the reactions. It is
the major contributor of the $S$-wave amplitude in the low-energy
region. In particular, in the region of $P_K\lesssim 300$ MeV/c,
$\Lambda(1405)S_{01}$ dominates the amplitudes, and contributions of
the other resonances are nearly invisible in the total cross
section. Around $P_K= 400$ MeV/c, the $\Lambda(1520)D_{03}$ is
responsible for the strong resonant peak in the total cross section.
Around $P_K=800$ MeV/c, the differential cross sections are
sensitive to the $\Lambda(1670)S_{01}$. In this energy region the
role of $\Lambda(1690)D_{03}$ is visible, but less important than
$\Lambda(1670)S_{01}$.

The non-resonant backgrounds, $u$- and $t$-channel, also play
important roles in the reaction. In the $t$-channel, the
$K^*$-exchange has larger cross sections than the $\kappa$. It
enhances the cross section obviously at the forward angles, and has
some destructive interferences at the backward angles. There can be
seen a small contribution of the $s$-channel $\Lambda$-pole, which
slightly enhances the cross section.

The $u$-channel significantly suppresses the differential cross
section at the forward angles, and produces the characteristic
backward enhancement. The significant contributions of the
$u$-channel agree with the results of U$\chi$PT
\cite{Oller:2006jw,Oller:2005ig}. In the quark model framework, the
$u$-channel allows transitions that the initial and final state
mesons can be coupled to the same quark or different quarks, while
the $s$-channel can only occur via transitions that the initial and
final state mesons are coupled to different quarks. This explains
the importance of the $u$-channel contributions. In comparison with
the U$\chi$PT, the agreement implies some similarity of the coupling
structure at leading order. For instance, the meson-quark couplings
in our model can be related to the meson-baryon couplings via
current conservation such as the recognition of the
Goldberger-Treiman relation~\cite{gt-relation}.

Our analysis suggests that there exist configuration mixings within
the $\Lambda(1405)S_{01}$ and $\Lambda(1670)S_{01}$ as admixtures of
the $[\textbf{70},^2\textbf{1},1/2]$ and
$[\textbf{70},^2\textbf{8},1/2]$ configurations. The
$\Lambda(1405)S_{01}$ is dominated by
$[\textbf{70},^2\textbf{1},1/2]$ (93\% or 57\%), and
$\Lambda(1670)S_{01}$ by $[\textbf{70},^2\textbf{8},1/2]$ (93\% or
57\%),  which is in agreement with the U$\chi$PT results
\cite{Jido:2003cb}. The $\Lambda(1520)D_{03}$ and
$\Lambda(1690)D_{03}$ are assigned as the
$[\textbf{70},^2\textbf{1},3/2]$ and
$[\textbf{70},^2\textbf{8},3/2]$, respectively. This prescription
indicates that the $\Lambda(1405)S_{01}$, $\Lambda(1520)D_{03}$ and
$\Lambda(1670)S_{01}$ still possess features of the traditional
3-quark states though they may also have some exotic properties
which are not sensitive to the measurement of the cross sections.
Experimental measurement of polarization observables may be more
selective for exposing their natures, especially for the
$\Lambda(1405)S_{01}$. Nevertheless, more accurate differential
cross sections in the low beam momentum region, e.g. $P_K=200\sim
500$ MeV/c, should also be useful.

For higher resonances, we expect more accurate data in the region of
$P_K=750\sim 900$ MeV/c can be useful for clarifying their
contributions and properties. With such data available, we can then
further study the role of $\Lambda(1670)S_{01}$,
$\Lambda(1690)D_{03}$ and the other higher $P$- and $F$-wave
resonances in the $n=2$ shell. The J-PARC facilities, which start to
run recently, will provide great opportunities for the study of the
hyperon spectrum in theory.

By comparing with approaches at hadronic level, so far we have not
yet included the coupled-channel dynamics. It would be interesting
and extremely useful to develop a coupled-channel calculation in our
framework for baryon resonance excitations in meson-nucleon
scattering and meson photoproduction. This would be a natural way of
restoring unitarity of the theory, and provide a microscopic
description for meson-baryon couplings. Nevertheless, with the
coupled-channel effects, one should be able to compare the quark
model form factors with those extracted from the hadronic models. We
wish to report the progress in the near future.


\section*{  Acknowledgements }

This work is supported, in part, by the National Natural Science
Foundation of China (Grants 10675131 and 10775145), Chinese Academy
of Sciences (KJCX3-SYW-N2), and the U.K. EPSRC (Grant No.
GR/S99433/01). The authors would like to thank useful discussions
with A. Sibirtsev, T.-S.H. Lee, and B.-S. Zou for very useful
discussions on relevant issues.


\end{document}